# Quality of Service with Bandwidth

Shivaji P. Mirashe, Dr. N.V. Kalyankar

**Abstract**— This paper deals with providing Quality of Service (QoS) over IP based networks. We are going to give a brief survey about this topic, and present our work at this area. There are many solutions of the problem, but the standardization of the methods is not finished yet. At the moment there are two kinds of approaches of the reservation problem. The distributed method handles the network nodes independently, and get the nodes making their own admittance decisions along the reservation path (i.e. Border Gateway Reservation Protocol BGRP. The centralized way -we discuss in details-, which collects the network nodes into domains, and handles them using a network manager. Generally there are two significant parts of the network management: intra domain, and inter-domain. This article focuses on making reservations over several domains, which is the part of the inter-domain functions.

**Index Terms**—Keywords are as Motivation and brief survey of providing QoS, IP QoS principles, IntServ, DiffServ, Bandwidth Broker, Inter-domain communication, Availability Information (AI) propagation, DS processing.

—————————— ◆ ——————————

## 1 INTRODUCTION

FIRST we give a short overview of the QoS providing over IP networks, and it's reason for the existence. In section two we discuss the principles need to be taken to provide QoS over IP networks. In the rest of this section we describe IntServ and DiffServ, and the QoS architecture using Bandwith Broker. At the end of this part we deal little with the ProFIS architecture. In the third chapter we introduce our inter-domain communication protocol for the ProFIS, and at the end of the document we are giving a summary of our work.

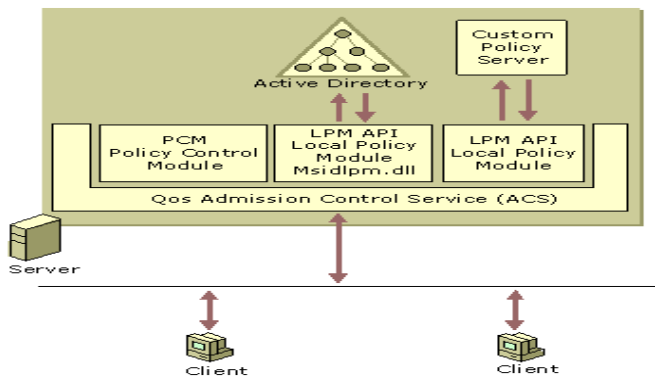

**Figure 1 QoS Admission Control Service**

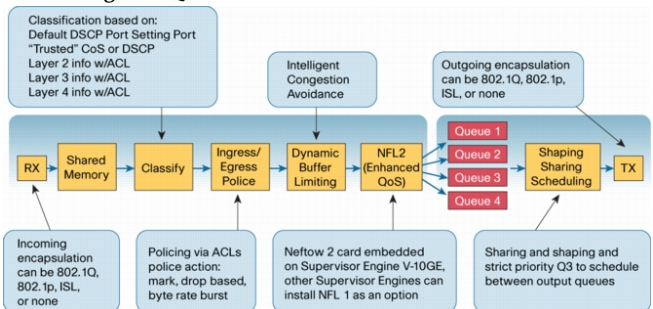

**Figure 2. Supervisor Engine QoS Processing**

## 2 MOTIVATION AND BRIEF SURVEY OF PROVIDING

————————————————
- *F.A. is Working as a Manager,IDC-Internet Data Center,Reliance Communications,New Mumbai.(Maharashtra) - (INDIA)*
- *S.A. Author Principal of Yeshwant Mahavidyalaya Nanded*

### QoS:-

Recent times the performance of the personal computers increases likewise the number of real-time Internet and multimedia applications. In case of these applications best effort traffic is not enough for satisfying the quality claim of the users. The best effort guarantee is an elementary provision of the Internet. The network elements try their best to deliver the packets to their destination without any bounds on delay, jitter, and latency, but they cannot give any guarantee for the delivery. These "guarantees" are not sufficient for i.e. a videoconference, because delay over a limit, or jitter can cut down or bust the interactivity and usability. The goal of the Internet Service Providers (ISPs) is to satisfy the quality demand of the customers and ensure the same sort of QoS and reliability over IP networks as in the circuit switched networks. By applying packet classification they can deliver different kind of services on the same link without the suffering of the important flows. The IP QoS is one of the most important research areas in our days. The development is driven by the increasing demands of the customers for service quality and reliability. Most of the technological challenges have been solved, now it is a matter of standardizing the technologies, and making the system scaleable. To solve the problem of scalability is one of the most important challenges because of the rapid growth of the modern Internet.

Main negative of the Public Switched Telephone Network (PSTN) that it can only deliver one kind of data. The IP technology is more flexible than any circuit's switched provision as it can carry different kind of traffic on the same link. The Internet is a complicated, heterogeneous system, which contains lot of different Autonomous System (AS) with different routing algorithms and different QoS technologies, applications. The number of the QoS architectures is high, but the interoperability and the standardizing are still not solved.

Since the actual Internet architecture does not provide mechanisms for resource management and isolation of the flows, all of the running services suffer in the congestion periods. Hence, in order to provide quality of service, an important step is to implement admission control mechanisms.

Other shortcoming of the current IP networks is that IP does not have the technical support for offering premium services. Each transmitted packet in the network is treated in the same way, the treating functions do not depend on the carried information of the packet, only on the destination address.

## 3 IP QoS PRINCIPLES:-

To solve the problems described above and provide QoS over IP networks, different principles have been developed:
- Packet classification is necessary to make the network capable to differentiate between the various categories of the traffic. The traffic must be classified based on its requested parameters like delay, jitter, bandwidth, price etc
- Isolation: scheduling and policing. The packets of the different categories must be scheduled different ways, and treated according to the policies. Each packet in the same QoS class is treated using the same method.
- High resource utilization is very important because of the economy view of the service provider.
- Call admission is required for accurate resource management in order to avoid congestion.

Although there are several approaches to the problem, two main QoS models can be considered for deployment: Integrated Services (IntServ) and Differentiated Services (DiffServ). These models are widely researched and accepted. The models or their combination seem to be good solution for providing QoS on the Internet.

As demonstrated in this document, the numbering for sections upper case Arabic numerals, then upper case Arabic numerals, separated by periods. Initial paragraphs after the section title are not indented. Only the initial, introductory paragraph has a drop cap.

## 4 INTSERV:-

First in 1993 Integrated Services was developed by Internet Engineering Task Force (IETF). In IntServ, a signaled QoS model is defined, where resource requirements are signaled from an endpoint, and the network device honors the signaling and reserves resources for flows. The protocol used for signaling here is Resource Reservation Protocol (RSVP) [5] . In addition to provide QoS per flow, in this model, admission control of traffic flows is available because of the inherent signaling ability.
**The RSVP has two main messages. The PATH and the RESV message.**
End stations, proxy devices, or voice gateways can initiate signaling by sending a PATH message with filter specification. The filter specification contains source, destination addresses and port numbers along with bandwidth requirement for the "flow", which can be defined as traffic going user to user, session to session, and gateway to gateway or proxy device to any other device in the network.

When the PATH message traverses the network along each hop the network device performs a check, initializes a state for the flow and forwards the Path message toward the destination. When the PATH message reaches the destination, the destination device responds with a RESV message including the bandwidth requirement for the flow. The RESV message traces its path back to the source. Along each hop the network device performs admission control. If there is enough available bandwidth, it admits the flow by assigning resources to it such as queues, weights, etc. and forwards the RESV message upstream toward the source. When the RESV message reaches the source, the signaling process is complete. When the service starts, network devices classify the traffic, recognize that it belongs to a reserved flow and put the packets into appropriate assigned queues so that the traffic gets the treatment it needs or signaled for.

Using IntServ and RSVP the flows must be administrated in every node along the path, and they are identified by their source and destination address. The number of the Internet hosts increases very quickly and ere now overshot the 120,000,000. Resource reservation schemes must scale well with the growth of the number of the hosts, because a router may be able to handle tens of thousands of reservations at same time. In case of the today's Internet the flow number can be a million in each router of the path, and such a big flow number debase the performance of the routers. Hence this method is not scalable for large networks because of the quick growth of the flows. One possibility to solve the scalability problem is that you do not distinguish between the packets, just the group of the packets. Other possibility is to group the nodes and so put hierarchy in the system. In the following chapter we examine the DiffServ model, which was developed driven by the demand of scalability.

## 5 DIFFSERV:-

DiffServ [1] as defined in the IETF RFC 2475 is a model that allows deployment of QoS in a simple fashion with network devices only handling traffic at an aggregate level rather than per flow.

The six most significant bits in the TOS byte of IP header is defined as DiffServ Code Point (DSCP). Packets are marked with a certain value depending on the type of treatment the packet must receive in the network device. Traffic is aggregated into traffic classes that require the same treatment and marked with a DSCP value in the TOS byte. DiffServ defines the DS domain, which is a continuos set of DiffServ capable nodes. The complexity of the network was moved to the edges of the domains. Filters are configured at the network ingress to identify the traffic and mark the traffic with the appropriate DSCP. The ingress routers are also responsible for policing and shaping. So inside these points a queue must be set up and a drop policy must be defined. Besides this, a policier-shaper must be configured and aggregated bandwidth must be allocated to the queue. The size of the DSCP is six bit, so it is able to manage up to 64 different behaviors. Hence the traffic of a DiffServ managed network from any source to any destination must fit into one of the 64 behaviors. In the DiffServ architecture each packet, which has the same DSCP, get the same treatment irrespective from the source and destination of the packet. So in a DiffServ node requires less entry inside a node than the flow entries in the same node in case of IntServ. It seems to be a good idea to manage the DS domains, because a domain manager is able to use the resources of the domain more effectively, and makes possible to serve the customer's claims. The domain manager of the DiffServ model is called Bandwidth Broker as introduced in RFC 2638.

## 6 BANDWIDTH BROKER (BB):-

Each domain has at least one BB, which is the manager of the domain's resources. The BB knows the topology, and has correct information about the currently reserved and free link capacities. The main functions of the BB are the following. Managing the resources of the own domain. These functions are the intra-domain functions. Another group of the BB functions is the inter-domain communication, it includes the communication of the availability information with the adjacent domain's Bandwidth Brokers, and the negotiation based on the received availability information.

Within the range of the intra-domain functions the BB manages the resources of the domain. If the BB receives a reservation from an end user, or an adjacent BB, checks the topology file. Inspects every link along the path, if the needed bandwidth fits in the unreserved capacities. If the claim can be satisfied the reservation can be admitted into the domain. After the decision was done the BB reserves the resources for the admitted reservation. Sets up the border routers. The main advantage of the DiffServ, and it makes it to a scalable architecture, that only the ingress routers need to be configured, because the routers inside the domain only forward the packets along the path towards the egress router according to the predefined per hop behavior. Because of the scalability of th earcitecture a good way to build a scalable network is to apply Bandwidth Broker managed DiffServ domains.

We consider two functions as as the inter-domain functions of the BB. Diffusing of the availability information towards the BBs of the adjacent DiffServ domain. The diffusion is important because the BB, and the customers of the domain have only information regarding to the available resources of the own domain. The diffusion is the way to get information about the available resources of the other domains. After receiving the availability information from the adjacent BB, the customer, or the BB can send a request to the BB about the resources he wants to reserve. If the request can be satisfied, the BB will get the resources from the adjacent domain. Consequently the negotiation is duable using this two method.

In the following section we describe Telia's DiffServ based BB managed architecture in few words, and after that present our work, which issued in a realized ProFis architecture with inter-domain communication functions.

## 7 THE PROFIS ARCHITECTURE:-

The ProFIS architecture is a system specification for providing end-to-end QoS over IP based networks. This concept uses the idea of the Bandwidth Broker, which means that the network domains are treated in a centralised way. The job of the BB is to handle the resources of a specified domain. The BB must determine whether the received bandwidth demands can be admitted to the network or not, and using this information it have to configure the border routers as well (intra-domain management). This approach can be prosperous, because equipment, which has knowledge about the whole domain, may handle the resources effectively. In the next part of this paper, we are going to describe an algorithm, which realises an inter-domain communication for the ProFIS architecture. The method is designed for especially this architecture, but we suppose that it contains useful parts for all the Bandwidth Broker architectures.

## 8 INTER-DOMAIN COMMUNICATION:-

First of all we have to mention, that this communication method is designed for handling aggregated demands. We presume, that the reservations are not made by edge users, but network providers, so they arrive periodically and not at a random time. This is a common case of reservations in backbone networks. These reservations have high bandwidth demands, and low bandwidth fluctuation. Considering this we can say that the state of the whole system scarcely differs from the previous state. We call this state of the system as constant state hereafter. The condition of the subsistence of the constant state is that the demands of the users are considerably constant, they send demand specifications periodically and there is no configuration change in the network.

At first, we will define a propagation algorithm. The goal of this is to provide the edge domains of the network, with proper information about the other available edge domains. We call this information as availability information (AI). One AI refers to one edge domain. This contents the name of the edge domain, the amount of the bandwidth, the average delay, the maximal delay, the delay, and the loss ratio. The ProFIS concept defines propagation steps for the event, when the constant state is set and some availability parameter, i.e. loss, delay, changes in the network. We made this theory complete by defining a method for the case when the configuration changes in the network i.e. we connect another transit domain to the network with another BB. The special of case of this, when the whole system stands up. We implemented a method to make the setting up of the system fast.

After this we are going to define a method, to reach and hold up the constant state of the system. The main idea is that the users send out demand specifications periodically at the edges of the network, and we generate automatic demand specifications in the internal network by aggregating the demands.

## 9 AVAILABILITY INFORMATION (AI) PROPAGATION:-

If these steps are done for each AI, all the BBs will possess exactly one AI for each edge domain in each service class and will be informed about the network parameter changes. It is important to mention that there can be loops in the network, and with a spreading method like this there, can be infinite loops in the propagation. Storing only one AI for each edge domain, which is the best, solves this problem here. If we received the best there will be no more AI change so the propagation stops.

**AI propagation for the standing up**

Now we are going to complete the method above with different propagation steps to make the system to handle network configuration change. We expect from this process all the BBs will have the proper AI even there is a configuration change in the network.

For doing this we define a special AI called NewAI and we make difference between the two AIs in the header field. The BB sends out NewAIs when it is standing up. The BB stands up in the following way:

At first it sends out the local AIs to the other BBs. We call an AI as local AI if the edge domain is directly connected to the domain, which is managed by the BB. The BB sends out the first AI as a NewAI, and sends out the other AIs as simple AI. By this step we reached that all the BBs will have information about the newly connected edge domains.

Finally the BB must gain the information about all the available edge domains, which are connected to the network. For solving this problem we use the fact that the other BBs are in the constant state so each BB possess exactly one AI for each edge domain. Now the BB is going to query the whole AI database from the neighboring BB. For this aim we use the special treatment of the NewSD. This method hardly differs from handling an AI. If the BB propagates the NewAI to the BB from which the NewAI is received than it will send the whole AI database to it. If the BB propagates to another direction, than makes a simple AI from the NewAI. Treating a NewAI is the following:

1. The BB determines whether the received AI is better then the stored.

2. The BB calculates the correct delay parameters and sends out a simple AI to all the BBs instead of the sender of the NewAI. If the propagation goes to the sender than the BB sends out the whole AI database to that BB.

Now the newly switched on BB will possess all the information about the available domains.

Using the idea of the NewAIs we can able to handle the problem of network errors, and network configuration changes. We can do it in using the "soft-state" principle. We provide the AIs with a validity time. The AIs have to be refreshed periodically else the referring edge domains are not available in the network.

Since the BBs store only one AI for each edge domain this communication method cannot realizes load balancing in the network, because the users are able to make reservations only for the route, which is evolved during the propagation. If a link is full on a route between two edge domains, the system will not be able to satisfy the demands even if there are free resources along other routes.

## 10 DS PROCESSING:-

This process realizes the bandwidth reservation mechanism. Let's see in a few steps how it works. We consider one cycle for one term to send the demand specifications.
1. The users send out the demand specifications from the edge domains. (At the beginning of the cycle)
2. Aggregating the DSs according to the destination information. Archiving the demands and sending out aggregated DSs automatically to the appropriate domains. (In the middle of the cycle)
3. On the basis of all the received demand specifications, sending the demands to the BB's intra-domain part, which will decide whether the demand can be satisfied or not, and configures the border routers. (At the end of the cycle)

This method handles the BBs independently, so there is no time synchronization between them. This means that different BBs achieve these points independently at a random time. This way can happen that a BB configures its domain just before it receives a demand specification. This demand can be satisfied only in the next term. There can be an unfortunate case, when an event like this happens along a whole route. Satisfying this kind of demand can suffer a huge delay. The principle of the ProFIS architecture, that the state of the system hardly differs from the previous state, covers this problem. We can say that in a constant state, it does not matter when the demands are satisfied, because there is a demand specification like that from previous states of the system. This way can happen that the user feels that his demand is served but it can occur that this is the affect of a DS, which has been sent few terms ago.

We made the step No. 3 in a simple way, which we give the reservations to the inter-domain communication in the sequence they arrived to the inter-domain module. This can cause that some demands can be fully achieved, but some demands cannot be satisfied because there are no free resources. The system informs the user about the unsuccessful reservation. There can be other solutions to make the sequence just by using game theoretical considerations.

## 11 CONCLUSION:-

As a result of this work we have given a suggestion how to realize End-to-End QoS over DiffServ capable network using Bandwidth Broker. It is important to mention that the introduced solution is only one of the several possibilities. The main advantages are that the number of the messages is low (each edge domain sends only one DS in any term in a constant state) and the system converges to a constant state rapidly. The system is scalable it consumes the network resources more efficient. Disadvantages are the sensitivity to the forecasts, which are not correct, other problem is that the DS message propagation time can be too long if the processing time points are quite different in each BB. Finally we notice that the inter domain protocol is recommended to be a standardized communication protocol. It is important because there can be several kind of implementations of these methods. We suggest using Extended Markup Language (XML) for this purpose. XML is a standardized language, it is easy to understand by human reading, and it is compatible with the World Wide Web. The legibility for humans is important because some parts of the DS processing cannot be automated. For example some decisions that are made by the machine are not overlaps with the financial considerations. For this reason, an interface has to be implemented for human interaction.

## 12 ACKNOWLEDGMENT:-

The authors wish to thank Miss. Suvran D. Alandkar (Mirash), Professor Shivaji Balaji Chavan (Yeshwant Mahavidyalaya Nanded), Mr. Aniket Despande (IBM), Mr. Satish Khadap (infovisionindia Consulting Services) & Mr.Satish D. Alandkar (BSNL-Pune). This work was supported in part by a grant from support to write this paper.

## 13 REFERENCES:-

**First Author:- Shivaji P. Mirashe**

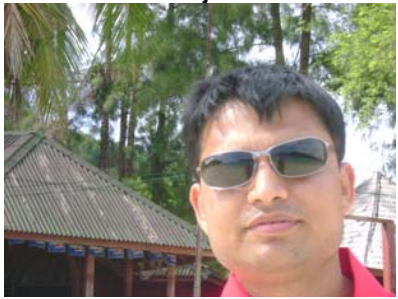

Myself Mr. Shivaji Pandurangrao Mirashe. I have completed MCA from S.R.T.M.U. Nanded (Maharashtra) Indian. I have got 8 Years experience in Information Technology. Currently I am working as a Manager in Information Security in Reliance Communication. Mumbai – Maharashtra (INDIA), doing the PHD at Yeshwant Mahavidyalaya Nanded, affiliated to S.R.T.M. University NANDED, Maharashtra, INDIA

**Paper Publish at different place & IEEE as below name,**

1) Firewall Penetration Testing Paper ID M575 – The 2nd International Conference on Computer modeling and simulation (iccms 2010)- http://iccms.org/
2) Peer-to-Peer Network Protocols ID H271 - The 2010 International Conference on Signal Acquisition and Processing (ICSAP 2010) - http://www.icsap.org/
3) Why We Need the Intrusion Detection Prevention Systems (IDPS) In IT Company – ID E447. - 2nd International Conference on Computer Engineering and Applications (ICCEA 2010) - http://www.iccea.org/
4) E-marketing, Unsolicited Commercial E-mail, and Legal Solutions – Emerging Trends in Computer Science, Communication & Information Technology (CSCIT2010) www.cscit2010.com
5) Saving the World Unsolicited Email Flow - Emerging Trends in Computer Science and Information Technology-2010
(For further information visit http://www.kkwagh.org/ETCSIT/ETCSIT10.html
6) Shivji Mirashe is a member of the IEEE and the IEEE Computer Society & International Association of Computer Science and Information Technology
IACSIT( Member NO. : 80337345) .

**Second Author:-**
**Namdeo V. Kalyankar:**
Dr. N.V. Kalyankar
Principal of Yeshwant Mahavidyalaya Nanded.
S.R.T.M.University
Nanded (Maharashtra) - (INDIA).
Completed M.Sc. Physics from B.A.M. University , Aurangabad. in 1980. in 1980 he joined as Lecturer in Department of Physics in yeshwant College,Nanded. In 1984 he completed his DHE. He Completed his Ph.D. from B.A.M.University in 1995. From 2003 he is working as Principal since 2003 to till date in Yeshwant college Nanded. He is also Research Guide for Computer Studies in S.R.T.M. University , Nanded. He is also worked on various bodies in S.R.T.M. University Nanded. He also published research papers in various international/ national journals. He is peer team member of NAAC (National Assessment and Accreditation Council)(India). He published a book entitled " DBMS Concept and programming in Foxpro". He also got "Best Principal" award from S.R.T.M. University, Nanded(India) in 2009. He is life member of Indian National Congress , Kolkata (India). He is also honored with "Fellowship of Linnean Society of London (F.L.S.)" on 11th Nov. 2009.